\documentclass[%
 reprint,
 amsmath,amssymb,
 aps,
]{revtex4-2}

\usepackage{graphicx}
\usepackage{dcolumn}
\usepackage{bm}

\usepackage{pdfpages}  
\usepackage{pgffor}

\makeatletter
\AtBeginDocument{\let\LS@rot\@undefined}
\makeatother

\usepackage[section]{placeins}   
\usepackage{float}               
\usepackage{dblfloatfix}         

\makeatletter
\def\fps@figure{!htbp}  
\def\fps@table{!htbp}
\makeatother

\begin{document}

\preprint{arxiv}

\title{On the Relationship and Distinction Between Atomic Density and Coordination Number in Describing Grain Boundaries}
\author{Reza Darvishi Kamachali$^{1,2,}$} 
\email{Corresponding author: reza.kamachali@bam.de (Reza Darvishi Kamachali)}
\author{Theophilus Wallis$^1$}

\affiliation{$^1$\textit{Federal Institute for Materials Research and Testing (BAM), 12205 Berlin, Germany}\\ $^2$Institute of Materials Physics, University of Münster, 48149 Münster, Germany}


\begin{abstract}
\noindent Crystal defects are often rationalized through broken-bond counting via the nearest neighbor coordination number. In this work, we highlight that this perspective overlooks intrinsic heterogeneities in interatomic spacing that decisively shape defect properties. We analyze excess free volume, energy, and entropy for a large set of BCC-Fe grain boundaries relaxed by molecular statics and demonstrate that an atomic-density field, as a systematically coarse-grained field variable, provides a more comprehensive descriptor.  Unlike coordination alone, the density field simultaneously captures bond depletion and spacing variations, thereby unifying structural and volumetric information. Our results establish density-based descriptors as principled surrogates for grain-boundary thermodynamics and kinetics, offer a direct bridge from atomistic data to mesoscale models, and motivate augmenting broken-bond rules in predictive theories of interfacial energetics, excess properties, segregation and phase behavior. 
\end{abstract}

\maketitle

One of the enduring concepts in modeling defects is the notion of broken bonds.
This model was originally articulated in the context of calculating surface energy and later extended to grain boundaries~\cite{read1950dislocation, hirth1982theory}.
This extension is well-justified, considering grain boundaries as two-dimensional crystallographic defects that disrupt the periodicity of the crystal lattice and involve a redistribution of atomic bonding environments.
Within this framework, the reduction in atomic coordination at the boundary is used as a proxy for the energetic cost associated with grain boundary formation, and therefore, the \textit{coordination number}---defined as the number of nearest-neighbor atoms---plays a central role. 

The coordination number serves as a structural descriptor in various atomistic and empirical models, such as the broken-bond rule for grain boundary energy \cite{du2005systematic, rohrer2011grain}, as well as in geometric and topological analyses of grain boundary networks \cite{mishin2010atomistic}.
It has also been widely applied in models of solute segregation, where it forms the basis for understanding the local chemical potential shifts near interfaces. 
For instance, in bond-breaking or bond-counting models of segregation \cite{seah1980adsorption, lejcek2010grain}, solute atoms preferentially occupy under-coordinated sites at the grain boundary to reduce the system’s energy. 
This principle underlies the McLean-type segregation isotherms \cite{mclean1957grain} and continues to inform modern statistical-mechanical and thermodynamic models of grain boundary segregation \cite{foiles1985calculation, murdoch2013grain}.

Despite its utility, the coordination number approach has inherent limitations: it discretely classifies neighbor counts but does not capture variations in interatomic spacing or local volume distortion.
This fundamental limitation motivates the exploration of coarse-grained field descriptors, such as atomic density, to generalize and extend beyond coordination-based models.

In a series of recent studies, it has been proposed \cite{darvishikamachali2020model} and extensively demonstrated \cite{darvishikamachali2020segregation, wang2021density, zhou2021spinodal, wang2021incorporating, wang2023calphad, wallis2023grain, kamachali2024giant} that atomic density, formulated as a coarse-grained field variable, serves as a unifying variable, enabling the seamless integration of atomistic simulations, CALPHAD thermodynamics, and phase-field models in studying grain boundaries.
The atomic density can be defined as
\begin{align}
    \rho_n(\vec{r}) &= \sum_I \delta(\vec{r} - R_I) ,
    \label{eq_delta_1}
\end{align}
which sums over all atoms in a given system, with spatial coordination $\vec{r}$ and the position vectors of the atoms $R_I$.
(The subscript $n$ in $\rho_n$ denotes the \emph{number} density, will be dropped after Eq. (\ref{eq_rho}) and should not be confused with the subscript later used in this monograph to indicate atomic coordination shells.)
The delta function can then be replaced by a three-dimensional Gaussian distribution function, thus giving
\begin{align}
    \rho_n(\vec{r}) &= \frac{1}{{(\beta\sqrt{2\pi})^3}} \sum_I e^{- \frac{\left(\vec{r}-R_I\right)^2}{2 \beta^2}} ,
    \label{eq_delta_2}
\end{align}
where $\beta$ is the smearing radius.
Considering then a reference defect-free bulk system, we obtain the corresponding bulk atomic density
\begin{align}
    \rho_n^B(\vec{r}) &= \frac{1}{(\beta\sqrt{2\pi})^3} \sum_I e^{- \frac{\left(\vec{r}-R_I^B\right)^2}{2 \beta^2}}
    \label{eq_delta_2_B}
\end{align}
Here the superscript $B$ refers to that defect-free \emph{bulk} system.
Given that an appropriate $\beta$ value is chosen, $\rho_n^B(\vec{r}) = \rho_n^B$ is a constant value within the bulk domain.
Thus, the final dimensionless atomic density field is defined as
\begin{align}
    \rho(\vec{r}) = \frac{\rho_n(\vec{r})}{\rho_n^B} \cdot
    \label{eq_rho}
\end{align}

Although Eqs. (\ref{eq_delta_1})--(\ref{eq_rho}) do not explicitly mention any relation with the coordination number, the sum in these equations which runs over all atoms in a given system inherently encodes information about the number of neighboring atoms.
To make this point clear, let us denote the number of atomic neighbors in the $n$-th coordination shell by $Z_n$, where $n = 1, 2, 3, \dots$ corresponds to the first-, second-, third-nearest neighbors, and so on.
The number of first-nearest neighbors, denoted as $Z_1$, is commonly referred to as the coordination number (a term originally coined from coordination chemistry).

In the reference bulk crystalline, the neighboring shells are defined based on the peaks in the radial distribution function, with each shell corresponding to a characteristic distance from a reference atom. 
Using Eq.~(\ref{eq_delta_2_B}) and the definitions outlined above, one can write the atomic density field in a defect-free bulk substance as
\begin{align}
    \rho^B_n(\vec{r}=R_I^B) &= \frac{1}{(\beta\sqrt{2\pi})^3} \sum_n Z_n^B \cdot e^{- \frac{\left(R_I^B-R_n^B \right)^2}{2 \beta^2}} \nonumber \\
    &= \phi_0^B + Z_1^B \cdot \phi_1^B + Z_2^B \cdot \phi_2^B + Z_3^B \cdot \phi_3^B + \dots ,
    \label{eq_delta_3}
\end{align}
where the summation is restructured from individual atoms $I$ to atoms in distinct coordination shells $n$, with $Z_n^B$ denoting the number of atoms in the $n$-th shell and $R_n^B$ denoting their equilibrium radial position (where $\vert R_I^B-R_n^B \vert$ is the shell radius).
For instance, in a perfect body-centered cubic (BCC) structure, the first (coordination) shell contains $Z_1^B = 8$ atoms, while the second shell contains $Z_2^B = 6$, and the third shell $Z_3^B = 12$, located at increasing radial distances.
Here 
\begin{align}
\phi_n^B = \frac{e^{- \frac{\left(R_I^B-R_n^B \right)^2}{2 \beta^2}}}{(\beta\sqrt{2\pi})^3}
\end{align}
represents the normalized Gaussian contribution from the $n$-th shell with respect to a reference atom at $R_I^B$.
$\phi_0^B$ corresponds to the center atom itself with $R_I^B = R_n^B$.
Thus, $\vert R_I^B-R_n^B \vert$ is the radius of $n$th shell.
This formulation implicitly assumes spherical symmetry and isotropic atomic environments, as appropriate for a perfect crystal.
The physical unit of $\phi_n$s is [length]$^{-3}$ and the inverse of it is a  \emph{coordination volume}.
Note that, for the sake of simplicity, Eq.~(\ref{eq_delta_3}) is written only at atomic positions $\vec{r}=R_I^B$, yet, the resulting value of the density field remains the same and a constant due to the translational symmetry of a defect-free crystal.

Referring now to disordered regions such as grain boundaries, the ideal shell structures become perturbed, resulting in local deviations in the number of neighboring atoms. This can be symbolically expressed as
\begin{align}
Z_n: Z_n^B \to \{Z_n\}^{GB}
\label{Eq_Zn}
\end{align}
where the bracket $\{\dots\}$ indicates a distribution and superscript GB denotes the grain boundary. 
This naturally induces variations in the coarse-grained atomic density field. 
Equation (\ref{eq_delta_3}) demonstrates that $\rho(\vec{r})$ directly reflects on the number of neighboring atoms.

Considering the first leading term, we can see that the atomic density is proportional to the coordination number:  $\rho^B_n(R_I) \approx Z_1^B \cdot \phi_1^B$.
However, it is readily clear that the \textit{interatomic spacing} between neighboring atoms must also adjust accordingly to accommodate local structural (and subsequently also chemical) inhomogeneities. 
In other words, not only does the discrete count of atoms in each coordination shell vary (Eq. (\ref{Eq_Zn})) but so too do the radial distances and the corresponding kernel weights
\begin{align}
R_n: R_n^B \to \{R_n\}^{GB},  \\
\phi_n: \phi_n^B \to \{\phi_n\}^{GB} \cdot
\end{align}
%

The above analyses suggest that the coordination number ($Z_1$), or even an extended set of higher-order neighbor counts ($Z_2$, $Z_3$ $\dots$), is insufficient to fully characterize defect structures. 
Instead, it is the combined information on both the number and spatial proximity of neighboring atoms that defines the structural nature of a defect.
These aspects both are captured by the atomic density field $\rho(\vec{r})$, such that regions with reduced coordination and/or expanded interatomic spacing—typically observed near disordered interfaces or defects manifest as local depletion in $\rho(\vec{r})$. 
Conversely, more densely coordinated and/or compressed regions correspond to local maxima. 
In contrast, coordination number metrics are insensitive to variations in interatomic spacing, inherently captured in the kernel contributions ($\phi_1$, $\phi_2$, $\dots$).

$\rho(\vec{r})$ encodes not only the coordination number but also the local volumetric distortion, providing a more comprehensive representation of the atomic environment than coordination numbers alone---even when considering multiple neighboring shells. 
This enriched descriptive power is crucial for accurately capturing grain boundary structure, long-range elastic fields, and excess volume, and is therefore essential for correctly computing the associated energy and entropy of the defect.

In the following we confront the two descriptors with data. 
We analyze a comprehensive set of $408$ distinct BCC-Fe grain boundaries (tilt, twist and mixed), generated and relaxed by molecular statics; protocol details are given elsewhere \cite{ratanaphan2015grain,wallis2025linking1}.

Recent analyses \cite{wallis2025linking1,wallis2025linking2} show that the \emph{grain boundary density}, defined as the average atomic density at the grain boundary plane,
\begin{align}
\rho^{GB} = \langle \rho(\vec{r}) \rangle_{\mathrm{at\ the\ GB\ plane}},
\end{align}
serves as a representative scalar quantity for the overall change in the atomic density field associated with a grain boundary. 

\newpage
\onecolumngrid 
\begin{figure*}[t]
    \centering
   \includegraphics[width=1\linewidth]{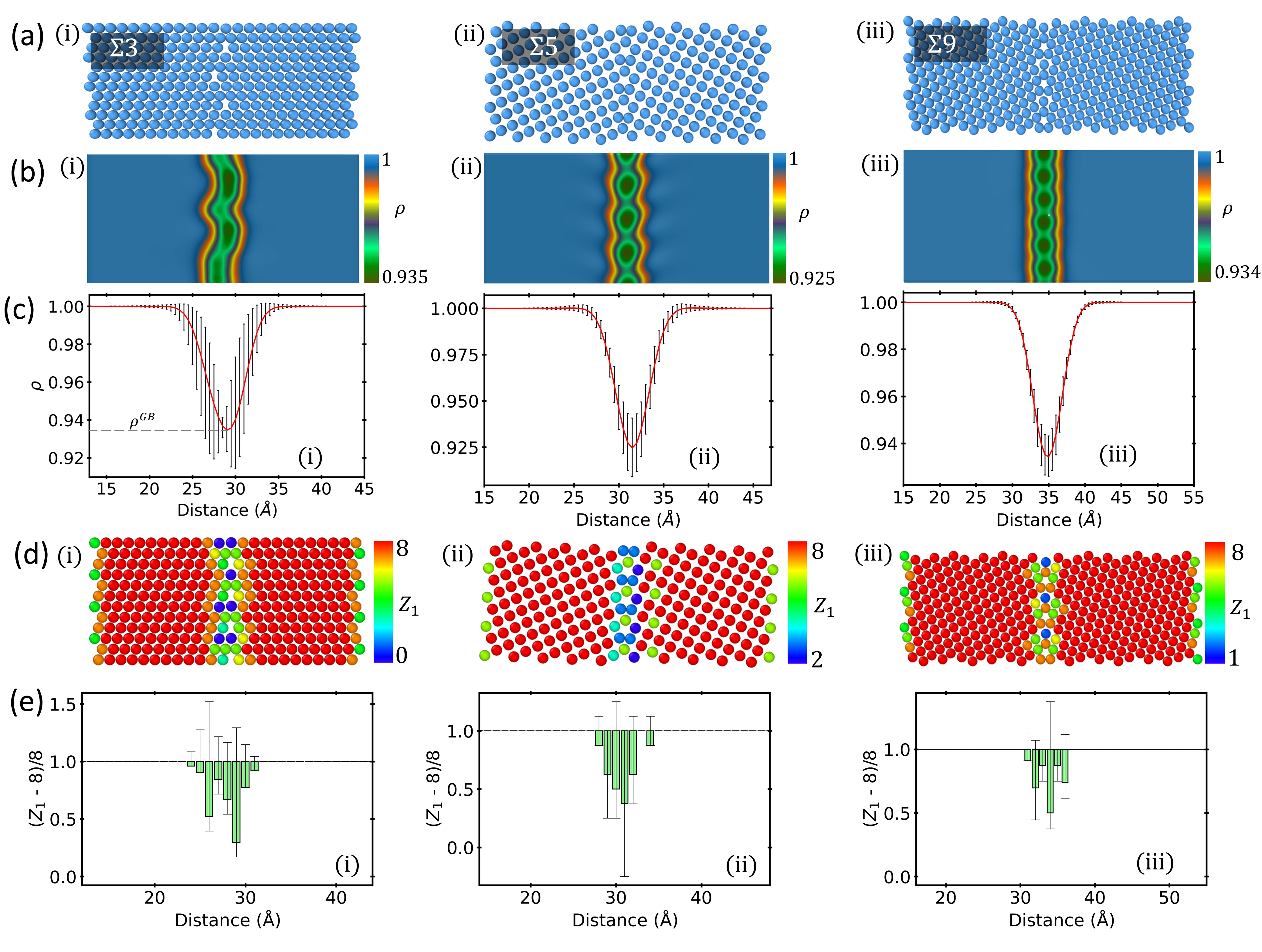}
   \caption{
    \textbf{Representative grain boundary examples.}
    (a) Atomistic structures for $\Sigma 3$, $\Sigma 5$, and $\Sigma 9$ boundaries; 
    (b) coarse-grained atomic density fields $\rho(\vec{r})$ showing depletion at the GB plane; 
    (c) density profiles $\rho(x)$ across the GB with $\rho^{GB}$ marked; 
    (d) atomic coloring by first coordination number $Z_1$; 
    (e) profiles of $(Z_1(x) - 8)/8$ across the GB with reduced values at the interface.
    The error bars in (c) and (e) show the range of the depicted quantity in the corresponding $yz$-slice.
    }
    \label{fig_gb_examples}
\end{figure*}

\twocolumngrid 

Analogous to $\rho^{GB}$, we can define the coordination number for every atom, average it at the GB plane
\begin{align}
Z_1^{GB} = \langle Z_1(R_I) \rangle_{\mathrm{at\ the\ GB\ plane}},
\end{align}
giving a similar dimensionless value $(8 - Z_1^{GB})/8$.

\begin{figure}[t]
    \centering
    \includegraphics[width=1.0\linewidth]{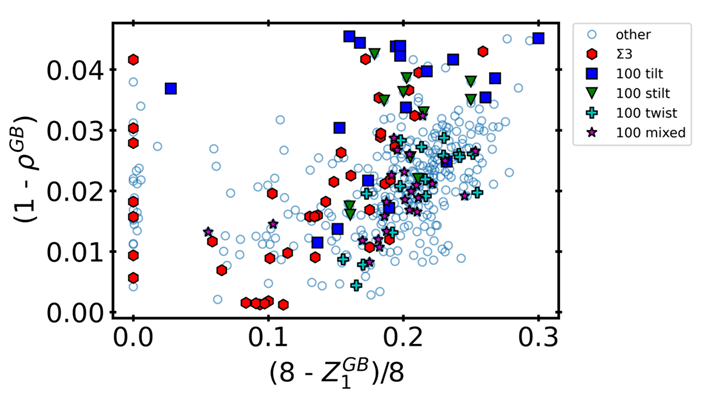}
\caption{
    \textbf{Correlation between density and coordination.}
    Scatter plot of density deficit $(1-\rho^{GB})$ versus coordination deficit $(8-Z_1^{GB})/8$ at the GB plane, across 408 BCC-Fe grain boundaries.
    Grain boundaries are classified by parent plane family, listed in the legend panel.
    This classification is the same in all following figures.
    The broad scatter demonstrates that $\rho^{GB}$ and $Z_1^{GB}$ are not equivalent descriptors.
    Notably, a sizable population has $Z_1^{GB} \approx Z_1^B(=8)$ while exhibiting a nonzero density deficit.
    }
    \label{fig_rho_z1}
\end{figure}

\begin{figure*}[t]
    \centering  \includegraphics[width=0.8\linewidth]{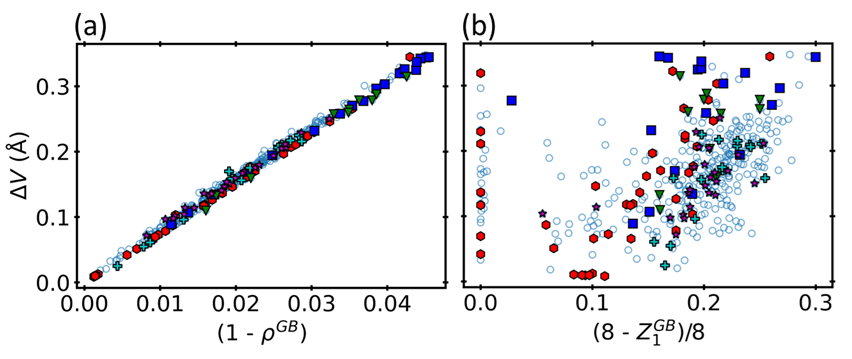}
    \caption{
    \textbf{GB excess free volume versus density and coordination.}
    (a) Excess free volume per unit area $\Delta V^{GB}$ collapses onto a near-perfect line with the density deficit $(1-\rho^{GB})$ across 408 BCC-Fe grain boundaries; 
    (b) $\Delta V^{GB}$ shows only a weak, scattered dependence on the coordination deficit $(8-Z_1^{GB})/8$, even for a given atomic plane family of grain boundaries. 
    }
    \label{fig_DeltaV_rho_z1}
\end{figure*}

\begin{figure*}[t]
    \centering
   \includegraphics[width=0.8\linewidth]{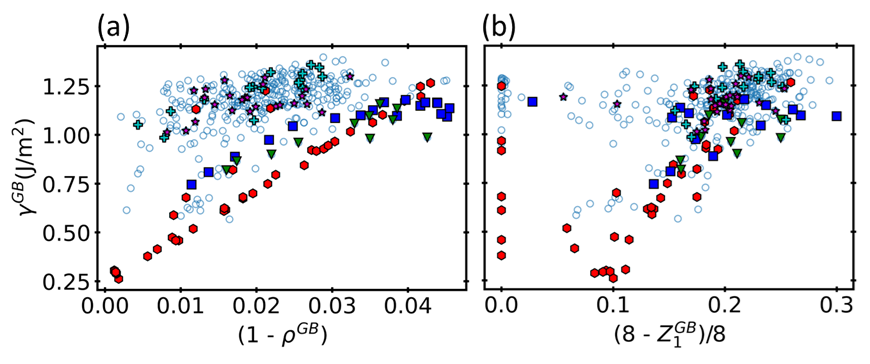}
    \caption{
    \textbf{GB energy versus density and coordination.}
    (a) Grain‐boundary energy $\gamma^{GB}$ increases systematically with the density deficit $(1-\rho^{GB})$ across 408 BCC-Fe grain boundaries;
    (b) in contrast, $\gamma^{GB}$ shows a weak, dispersed dependence on the coordination deficit $(8 - Z_1^{GB})/8$, including many cases with $(8 - Z_1^{GB})/8 \approx 0$ but finite energy.
    Grain boundaries are classified by parent plane family.
    }
    \label{fig_gamma_rho_z1}
\end{figure*}

\begin{figure*}[t]
    \centering
   \includegraphics[width=0.8\linewidth]{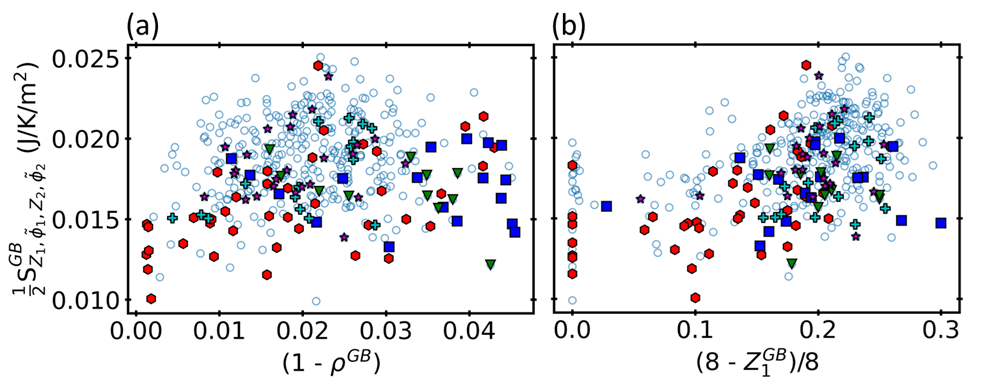}
   \caption{
    \textbf{Configurational entropy versus density and coordination.}
    (a) The sum configurational entropy per unit area $\tfrac{1}{2}\,S^{GB}_{Z_1,\tilde\phi_1,Z_2,\tilde\phi_2}$ increases systematically with the density deficit $(1-\rho^{GB})$, reflecting growth of combined topological-distantial disorder with density depletion;
    (b) the same quantity versus the coordination deficit $(8 - Z_1^{GB})/8$ exhibits only a loose relation with pronounced scatter, including many grain boundaries with $(8 - Z_1^{GB})/8 \approx 0$ spanning nearly the full entropy range.
    }
    \label{fig_entropy_4sources}
\end{figure*}

\begin{figure}[t] 
    \centering 
    \includegraphics[width=1.0\linewidth]{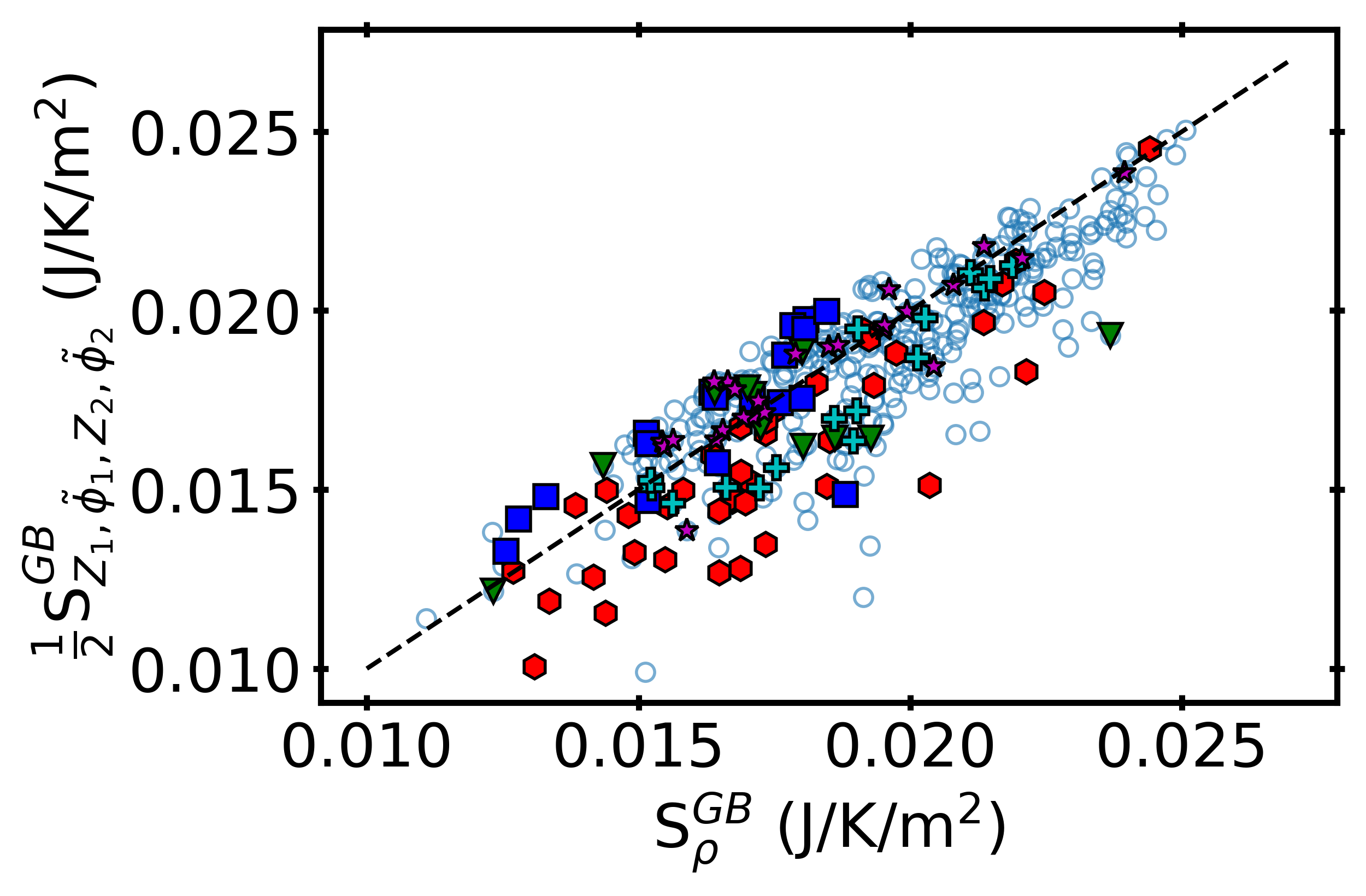} 
    \caption{
    \textbf{Equivalence of density and sum entropy.} 
    Areal configurational entropy from the density field, $S_\rho^{GB}$, plotted against the sum entropy $\tfrac{1}{2}\,S^{GB}_{Z_1,\tilde\phi_1,Z_2,\tilde\phi_2}$; data cluster tightly about the $y{=}x$ line (dashed), indicating a near-unity slope and small intercept across 408 BCC-Fe grain boundaries, thereby validating the sum relation and underscoring the information captured by the atomic-density field beyond coordination counts. 
    } 
    \label{fig_entropy_rho} 
\end{figure}

For each boundary we evaluate the \emph{density deficit} $(1-\rho^{GB})$ and the \emph{coordination deficit} $(8-Z_1^{GB})/8$, see Methods. 
These are then compared against three independent GB properties: 
the excess free volume per unit area $\Delta V$, the GB energy $\gamma^{GB}$, and two measures of configurational entropy per unit area—one obtained from the atomic-density field $S_\rho^{GB}$ and the other from shell descriptors, the \emph{sum entropy} $\tfrac{1}{2}\,S^{GB}_{Z_1,\tilde\phi_1,Z_2,\tilde\phi_2}$. 
Throughout, boundaries are grouped by parent-plane family for reference. 
We begin with representative examples (Fig.~\ref{fig_gb_examples}), then examine the joint variation of the two deficits (Fig.~\ref{fig_rho_z1}), followed by their relations to $\Delta V$ (Fig.~\ref{fig_DeltaV_rho_z1}) and $\gamma^{GB}$ (Fig.~\ref{fig_gamma_rho_z1}). Finally, we consider the entropic content (Fig.~\ref{fig_entropy_4sources}) and demonstrate the equivalence between $S_\rho^{GB}$ and the sum entropy (Fig.~\ref{fig_entropy_rho}).

Figure \ref{fig_gb_examples} shows representative grain-boundary structures alongside the coarse-grained density fields and coordination maps, indicating that the planar-averaged density $\rho(x)$ exhibits a clear depletion at the interface that robustly locates the GB, while the $Z_1$ maps register coordination loss only where it occurs. 
All three $\Sigma 3$, $\Sigma 5$ and $\Sigma 9$ boundaries show coincident reductions in both $\rho$ and $Z_1$, however, it is clear that the density field results in a consistent definition of the grain boundary region for all three examples, whereas the variations in coordination number are rather case-specific.
Figure \ref{fig_rho_z1} shows the scatter of the density deficit $(1-\rho^{GB})$ against the coordination deficit $(8-Z_1^{GB})/8$ at the GB plane, for all 408 GBs, indicating a loose relation between the two descriptors: 
many boundaries exhibit $(8-Z_1^{GB})/8 \approx 0$ while maintaining a finite density deficit, and for a given $(1-\rho^{GB})$ the corresponding $(8-Z_1^{GB})/8$ spans a wide range.
This non-equivalence anticipates the trends in Figs. \ref{fig_DeltaV_rho_z1}--\ref{fig_gamma_rho_z1}, where $(1-\rho^{GB})$ proves the more predictive scalar for $\Delta V$ and $\gamma^{GB}$.
Grain boundaries are classified and colored based on their parent atomic planes.

Recently we have demonstrated that GB atomic density reveals a remarkable correlation with the excess free volume \cite{wallis2025linking1}.
This is a nontrivial result as the GB density is the single (minimum) average density value at the grain boundary plane whereas the excess free volume is an integral property of the grain boundary. 
Figure \ref{fig_DeltaV_rho_z1} contrasts the excess free volume per unit area, $\Delta V$, with the density deficit $(1-\rho^{GB})$ and the coordination deficit $(8-Z_1^{GB})/8$. 
Panel (a) shows that $\Delta V$ collapses onto an almost perfectly linear trend with $(1-\rho^{GB})$ across all $408$ boundaries, with only minor family-dependent spread, i.e.\ the single GB-plane quantity $\rho^{GB}$ reliably predicts the integral excess volume. 
By contrast, panel (b) exhibits a weak, highly scattered relation with $(8-Z_1^{GB})/8$, including a sizable vertical band at $(8-Z_1^{GB})/8\!\approx\!0$ where $\Delta V$ spans a wide range. 
These boundaries --with nonzero excess volume but no coordination deficit--demonstrate that variations in interatomic spacing and local dilation, captured by the density field, are essential for excess volume, whereas coordination counts alone systematically miss this contribution.

Figure \ref{fig_gamma_rho_z1} summarizes the energetic trends: 
$\gamma^{GB}$ increases monotonically with the density deficit $(1-\rho^{GB})$ with a comparatively clear family-dependent spread, whereas its dependence on the coordination deficit $(8-Z_1^{GB})/8$ is weak and highly dispersed. 
Notably, many boundaries with $(8-Z_1^{GB})/8 \approx 0$ still carry substantial energy, demonstrating that elastic/dilatational contributions, captured by the density field, set the energetic scale that broken-bond counts alone cannot capture.

In order to quantify and discern the significance of the changes in interatomic spacing, we further our analyses by computing the entropy due to the disorder within a given system induced by a grain boundary.
In principle, the relevant measure is the Shannon entropy, which arises from the combined topological (bond depletion) and distantial (variations in interatomic spacing) disorder within the grain boundary.
These two aspects are respectively captured in Eq. (\ref{eq_delta_3}) where both coordination numbers and volumes are listed.
Clearly, other degrees of disturbance can still be imagined, e.g., the shape of closed volumes around every atom, which we omit from our entropy calculations.
For a grain boundary region, truncating the atomic density formula to the first two terms, we can write
\begin{align}
    \rho_n(\vec{r}=R_I) &\approx Z_1(R_I) \cdot \tilde\phi_1(R_I) + Z_2(R_I) \cdot \tilde\phi_2(R_I)
    \label{eq_rhon_truc}
\end{align}
where $\rho_n(\vec{r}=R_I)$ is computed based on Eqs. (\ref{eq_delta_1})--(\ref{eq_rho}) and the right-hand side can be obtained for each atom such that $\tilde\phi_1$ and $\tilde\phi_2$ are 
\begin{align}
    \tilde\phi_{1(or\ 2)}(R_I) = \frac{e^{-\frac{\left(R_I-\tilde{R}_{1 (or\ 2)}\right)^2}{2\beta^2}}}{(\beta\sqrt{2\pi})^3}
    \label{eq_effective_phi}
\end{align}
with $\tilde{R}_{1}$ and $\tilde{R}_{2}$ being effective radii computed based on the respective effective volume of the first and second coordination shell around a given atom.

We compute the entropy in two ways: once from the spatial distribution of the atomic density field across the grain boundary, $S_\rho^{GB}$, and once from per atom distributions of the descriptors, $S_q^{GB}$, with $q$: $Z_1,\,Z_2,\,\tilde\phi_1,\,\tilde\phi_2$.
See Methods for the precise definitions and procedure. 
From the descriptor entropies, the \emph{sum entropy} reads
\begin{align}
\tfrac{1}{2} S^{GB}_{Z_1,\tilde\phi_1,Z_2,\tilde\phi_2} = \frac{1}{2} \left( S^{GB}_{Z1} + S^{GB}_{\tilde\phi_1} + S^{GB}_{Z2} + S^{GB}_{\tilde\phi_2} \right)
\end{align}
See Methods section for details.
One of the main results of our study is the numerical equivalence of the density-based entropy and the sum entropy, $S_\rho^{GB} \approx \tfrac{1}{2} S^{GB}_{Z_1,\tilde\phi_1,Z_2,\tilde\phi_2}$, thereby confirming that the atomic-density field compactly captures the joint topological--distantial disorder encoded by the shell descriptors.

Figure \ref{fig_entropy_4sources} presents the sum entropy $\tfrac{1}{2}\,S^{GB}_{Z_1,\tilde\phi_1,Z_2,\tilde\phi_2}$ per unit area against the two GB-plane descriptors. 
Panel (a) in Fig. \ref{fig_entropy_4sources}  shows a monotonic increase of the sum entropy with the density deficit $(1-\rho^{GB})$, consistent with the notion that combined topological-distantial disorder grows as the atomic density depletes at the interface.

\newpage
\onecolumngrid

\twocolumngrid

Panel (b) in Fig. \ref{fig_entropy_4sources} contrasts this with a loose, heteroscedastic relation versus the coordination deficit $(8-Z_1^{GB})/8$, including many boundaries with no coordination deficit that nevertheless span nearly the full entropy range. 
In particular, this decoupling confirms that coordination counts alone underestimate the configurational disorder when spacing fluctuations dominate, whereas the density field --by construction-- captures both ingredients.

Figure \ref{fig_entropy_rho} then compares the two entropy constructions directly.
We find that $S_\rho^{GB}$ aligns tightly with $\tfrac{1}{2}\,S^{GB}_{Z_1,\tilde\phi_1,Z_2,\tilde\phi_2}$ along the $y{=}x$ line, with a near-unity slope and a small intercept across all families. 
This validates the \emph{sum relation} and shows that $S_\rho^{GB}$ is a compact, lossless proxy for the joint shell-descriptor information: 
the atomic density field encodes the combined effect of neighbor multiplicities and interatomic spacing, whereas $Z_1$ alone does not.

A particular point of discussion reflects in our results is about a large group of grain boundaries, of different classes, that have $Z_1^{GB} \approx 8$.
For twin and some special boundaries where the atomic coordination numbers at the grain boundaries remain the same as in the bulk lattice, DFT calculations show that segregation can still happen and the electronic contributions to the segregation energy can be significant, indicating that the chemical bonding between solute and host atoms is quite different from that of grain interior \cite{hu2020solute}.
This is precisely what is expected for the vertical population at $(8 - Z_1^{GB})/8 \approx 0$ observed here.

\newpage
\onecolumngrid

\twocolumngrid

For the same populations we have $(1 - \rho^{GB}) > 0$ and a finite $\Delta V$ and $\gamma^{GB}$ that can explain subsequent segregation behavior.
In contrast, we find that the density framework is agnostic to grain boundary character and thus captures features that are \emph{generic} across classes. 
Considering low-angle boundaries, where the structure is well-described as an array of dislocations (Read-Shockley type) with long-range elastic fields, the local coordination remains essentially bulk-like over most sites, yet the dilatational/compressional strains induce a measurable depletion in the coarse-grained density at the grain boundary plane. 
Conversely, in more disordered or high-angle boundaries, where substantial local volume changes and topology disruptions occur, both $Z_1$ and $\rho$ vary, but the density field integrates spacing heterogeneity and excess volume more faithfully, yielding the tight trends with $\Delta V$, $\gamma^{GB}$, and the growth of configurational entropy. 

Taken together, the results show that $\rho(\vec r)$ unifies long-range elastic effects (with minimal coordination change) and strongly disordered cores (with both topology and spacing changes), thereby providing a single state variable that is predictive across the full GB spectrum.
These aspects motivate that the atomic density field parameter can be very useful in studying the mechanical \cite{frolov2012thermodynamicsI,frolov2012thermodynamicsII,dehm2022implication} or electrical \cite{bishara2020approaches} response of grain boundaries, in the same spirit of the grain boundary excess volume.
Another remarkable aspect $\rho(\vec r)$ and $\rho^{GB}$ is their potential to link with the experimental measurements:
Quantitative routes to measure local atomic density at (deformation) interfaces are developed by combining HAADF-STEM with EELS thickness calibration and 4D-STEM with machine learning techniques \cite{Rosner2014Ultramicroscopy,Schmidt2015PRL,buranova2016,cheng2021acsNano}.
Independent radiotracer experiments revealed orders-of-magnitude enhancement of atomic diffusion inside free-volume-rich short-circuit paths \cite{Bokeloh2011PRL} which can also be linked with the atomic density field.

To summarize, the above analyses show that a coarse-grained atomic density field, $\rho(\vec r)$, captures the essential grain boundary physics and more completely than coordination number. 
While $Z_1$ registers broken bonds, $\rho$ embeds both topology (neighbor multiplicity) and spacing (local dilation/compression). 
This distinction is not semantic: 
across $408$ BCC-Fe boundaries the density deficit $(1-\rho^{GB})$ collapses the excess free volume $\Delta V$ onto an almost perfectly linear trend and organizes the GB energy $\gamma^{GB}$ with modest family dependent spread, whereas $(8-Z_1^{GB})/8$ shows broad, heteroscedastic scatter and admits many boundaries with \emph{no} coordination deficit but finite $\Delta V$ and $\gamma^{GB}$. 
These cases diagnose elastic/dilatational contributions that are invisible to broken-bond counting but are, by construction, encoded in $\rho$.

A second conceptual outcome concerns configurational disorder. 
We show that the density-based entropy $S_\rho^{GB}$ is able to track the sum entropy $\tfrac12\big(S_{Z_1}^{GB}+S_{\tilde\phi_1}^{GB}+S_{Z_2}^{GB}+S_{\tilde\phi_2}^{GB}\big)$,  nearly one-to-one. 
This numerical equivalence indicates that $\rho$ is a compact, effectively lossless proxy for the joint topological-distantial disorder. 
In information terms, the strong in-shell correlations (between $Z_s$ and $\tilde\phi_s$) imply substantial redundancy; 
the observed one-half prefactor reflects an effective two-mode structure dominated by the first two shells. 
Thus, $\rho$ summarizes what matters for GB thermodynamics without proliferating descriptors.

Practically, these results argue for augmenting broken-bond rules with density-based descriptors in models of GB energetics, excess properties, and segregation thermodynamics. 
Because $\rho$ is a field, it interfaces naturally with continuum treatments (e.g., CALPHAD/phase-field) and can carry elastic fields and volumetric excess in a consistent way. 
The subsumption of coordination number, interatomic spacing, excess volume and isotropic elasticity within a single density-based framework suggests that $\rho(\vec r)$ is a convenient surrogate and a candidate descriptor of crystalline disorder. 
Limitations remain: 
our dataset is BCC-Fe at $0$\,K (molecular statics), the coarse-graining length $\beta$ sets resolution and must be chosen judiciously, and the entropies used here are information-theoretic measures rather than full thermodynamic entropies. 
Nonetheless, the robustness of the trends across orientations and families suggests that the density framework is structurally generic. 
Expanding the current comparative analyses to other lattice structures, finite temperature, multi-component alloys with segregation, and kinetic phenomena (faceting, complexion transitions) should be particularly informative, with $\rho$ providing a unifying state variable across atomistic and mesoscale descriptions.
Also revisiting experimental procedures to study density deficits and excess volumes with diffraction, positron annihilation, high-resolution microscopy or even daring atom probe tomography measurements would provide decisive advancement in closing the loop between atomistics, mesoscale models, and experiment in studying microstructure defects.

\section*{Methods}
\emph{Atomistic Simulation Dataset}:
We analyze a set of $408$ distinct grain boundaries (GBs) in BCC-Fe generated and minimized following established protocols \cite{ratanaphan2015grain,wallis2025linking1}. 
Bicrystals are constructed with periodicity parallel to the GB and sufficient padding normal to the interface to suppress image interactions. 
After rigid-body scans and in-plane relaxations, structures are relaxed to a force and energy tolerance by conjugate-gradient minimization. 
Grain boundary energies and excess free volumes are computed using the atomistic simulations, for each grain boundary and both are per unit area.
All coarse-graining and further analyses reported below are performed on the minimized configurations. 
Further details of structure generation and parameterization as well as on the methods of coarse-graining atomic density field are provided in \cite{ratanaphan2015grain,wallis2025linking1}; 
for the present study, we only summarize the definitions needed in the body of text.
For visualization we group grain boundaries by parent plane families and special boundaries (e.g., $\Sigma 3$) following \cite{wallis2025linking1}. 
Symbols and colors in the figures map consistently across panels; all scalars shown with appropriate units, or are dimensionless (for the atomic density and coordination number deficit at the GB plane).

\emph{Atomic density and coordination number}: 
In each system and at the position of each atom, we compute atomic density value ($\rho(\vec{r} = R_I)$), coordination number ($Z_1 (R_I)$) and 2nd coordination number ($Z_2 (R_I)$).
Note the $\rho(\vec{r})$ can be computed for every position in the system.
The coarse-grained atomic density computation is of course sensitive to coarse-graining length (smearing radius $\beta$ in the Gaussian distribution).
This sensitivity is thoroughly discussed in \cite{wallis2025linking1}.
We use $\beta = 2.4 r_{Fe}$ with $r_{Fe} = 1.26$ {\AA} being the atomic radius of iron.
The GB plane is defined as the location of the minimum of the planar-averaged density along the interface normal. 
We compute
\begin{align}
\langle \rho \rangle(x) = \frac{1}{A}\int_A \rho(\vec{r})\,dy\,dz,
\end{align}
identify $x_\mathrm{GB}=\arg\min_x \langle\rho\rangle(x)$, and evaluate GB-plane averages over a narrow slab of thickness $w$ centered at $x_\mathrm{GB}$,
\begin{align}
\rho^{GB} = \frac{1}{Aw}\int_{x_\mathrm{GB}-w/2}^{x_\mathrm{GB}+w/2}\!\!\int_A \rho(\vec{r})\,dx\,dy\,dz .
\end{align}
where $w = 0.2$ \AA \ is the bin size in saving the continuous coarse-grained atomic density field.
All GB-resolved scalar quantities below use the same slab.

Nearest-neighbor coordination numbers are obtained from a bulk-referenced cutoff at the first minimum of the bulk radial distribution function, yielding $Z_1(R_I)$ at each atom position.
We form the GB-plane average
\begin{align}
Z_1^{GB} = \langle Z_1\rangle_{\mathrm{GB\ slab}}, \qquad
\delta Z_1^{GB} = 1 - \frac{Z_1^{GB}}{Z_1^B}.
\end{align}
with $Z_1^B = 8$.

To quantify spacing effects we use Eqs. (\ref{eq_rhon_truc}) and (\ref{eq_effective_phi}) effective shell weights $\tilde\phi_n$ are defined.
First, we compute the local shell volume for $Z_2$, that is the Voronoi volume for BCC structure, to compute an equivalent radius $\tilde R_2$ and evaluating the same Gaussian kernel $\tilde\phi_2$.
Using above computed data and Eqs. (\ref{eq_rhon_truc}) we compute $\tilde\phi_1 \approx \frac{\rho_n - Z_2 \tilde\phi_2}{Z_1}$ for each atom.
As $\tilde R_n$ is computed from the spatial coordinates $\vec{r}$ of neighbors (not from any topological count), using mid-shell boundaries between consecutive neighbor distances to define the shell volume. 
The pair $(Z_n,\tilde\phi_n)$ thus separates topology (counts) and distance (spacing).

\emph{Configurational Shannon entropy from atomic density and from Shell descriptors}:
We compute configurational Shannon entropies from spatial per-atom distributions that are obtained from our analyses. 
Note that this is not the same thing as thermodynamic entropy:
Shannon entropy is the entropy merely due to the disorder in the given system.
In our atomistic simulation set-up the only source of disorder is the (single) grain boundary.
We thus can compute this quantity per unit area of the grain boundary. 
For every property $q$,
\begin{align}
p_q = \frac{|q|}{\sum_J |q|}, \qquad
S_q^0(J) = -k_B \sum_J p_{q}\ln p_{q},
\end{align}
where $q \in  \big(\rho_n, \ \rho_n^B,\ Z_1,\ Z_1^B,\ \tilde\phi_1,\ \tilde\phi_1^B ,\ Z_2,\ Z_2^B ,\ \tilde\phi_2 ,\ \tilde\phi_2^B \big)$ and $J$ indicates the number of atoms in the whole system ($J = I$) or a bulk subsystem ($J = I^B$).
Precisely, $I$ considers ALL atoms and $I^B$ indicates Bulk atoms away from the GB plane (10 \AA) and also non-periodic boundaries along the $x$ axis (removed by indexing those atoms).
Thus, we compute the grain boundary excess entropy
\begin{align}
S_q^{GB} &= \frac{N^{GB}}{A} \left( S_q^0(\mathrm{ALL}) - S_q^0(\mathrm{Bulk}) \right) \nonumber \\
&= \frac{N^{GB}}{A} \left(S_q^0(J = I) - S_q^0(J = I^B) \right).
\end{align}
For example, $S_\rho^{GB} = \frac{N^{GB}}{A} \left( S_\rho^0 - S_{\rho^B}^0 \right)$ where $N^{GB}$ and $A$ are the number of atoms and area of the grain boundary.

\emph{Sum entropy and correlation analysis}:
To rationalize the empirical relation $S_\rho^{GB} \approx \tfrac12\sum_X S_X$, with $X=(Z_1,\tilde\phi_1,Z_2,\tilde\phi_2)$, we analyze the Pearson cross-correlations among six descriptor pairs $r(Z_1,Z_2)$, $r(Z_1,\tilde\phi_1)$, $r(Z_1,\tilde\phi_2)$, $r(Z_2,\tilde\phi_1)$, $r(Z_2,\tilde\phi_2)$, $r(\tilde\phi_1,\tilde\phi_2)$ per each grain boundary,
\begin{align}
r(x,y) = \frac{\frac{1}{N}\sum_{I-I^B} (x_i-\bar x)(y_i-\bar y)}{\sqrt{\frac{1}{N}\sum_{I-I^B} (x_i-\bar x)^2} \,\sqrt{\frac{1}{N}\sum_{I-I^B} (y_i-\bar y)^2}},
\end{align}
with $N$ the number of atoms in the corresponding $I - I^B$ region.

The resulting cross-correlations reveal a clear structure: 
We observe that most notably $(Z_1, \tilde\phi_1)$ and $(Z_2,\tilde\phi_2)$ are dominated by strong correlations. 
These results indicate that the \emph{count} and \emph{spacing} in each shell have high overlaps in their topological contribution.
At the same time we observe that only first shell is not enough to match the accuracy obtainable by the atomic density.
Taken together, these patterns indicate that the information carried by $(Z_1,\tilde\phi_1,Z_2,\tilde\phi_2)$ are contributing but with $\simeq$ 50\% redundancy. 
The statistical balance of strong positive and strong negative correlations gives
\begin{align}
S_\rho^{GB} \approx \tfrac{1}{2} S^{GB}_{Z_1,\tilde\phi_1,Z_2,\tilde\phi_2} = \frac{1}{2} \sum_{X \in (Z_1,\tilde\phi_1,Z_2,\tilde\phi_2)} S_X^{GB},
\end{align}
with the factor of $1/2$ reflecting the effective redundancy among descriptors revealed by their correlation structure.
Extensive details on all entropy-related calculations will be soon added as Supplementary Material.


\begin{thebibliography}{31}%
\makeatletter
\providecommand \@ifxundefined [1]{%
 \@ifx{#1\undefined}
}%
\providecommand \@ifnum [1]{%
 \ifnum #1\expandafter \@firstoftwo
 \else \expandafter \@secondoftwo
 \fi
}%
\providecommand \@ifx [1]{%
 \ifx #1\expandafter \@firstoftwo
 \else \expandafter \@secondoftwo
 \fi
}%
\providecommand \natexlab [1]{#1}%
\providecommand \enquote  [1]{``#1''}%
\providecommand \bibnamefont  [1]{#1}%
\providecommand \bibfnamefont [1]{#1}%
\providecommand \citenamefont [1]{#1}%
\providecommand \href@noop [0]{\@secondoftwo}%
\providecommand \href [0]{\begingroup \@sanitize@url \@href}%
\providecommand \@href[1]{\@@startlink{#1}\@@href}%
\providecommand \@@href[1]{\endgroup#1\@@endlink}%
\providecommand \@sanitize@url [0]{\catcode `\\12\catcode `\$12\catcode
  `\&12\catcode `\#12\catcode `\^12\catcode `\_12\catcode `\%12\relax}%
\providecommand \@@startlink[1]{}%
\providecommand \@@endlink[0]{}%
\providecommand \url  [0]{\begingroup\@sanitize@url \@url }%
\providecommand \@url [1]{\endgroup\@href {#1}{\urlprefix }}%
\providecommand \urlprefix  [0]{URL }%
\providecommand \Eprint [0]{\href }%
\providecommand \doibase [0]{https://doi.org/}%
\providecommand \selectlanguage [0]{\@gobble}%
\providecommand \bibinfo  [0]{\@secondoftwo}%
\providecommand \bibfield  [0]{\@secondoftwo}%
\providecommand \translation [1]{[#1]}%
\providecommand \BibitemOpen [0]{}%
\providecommand \bibitemStop [0]{}%
\providecommand \bibitemNoStop [0]{.\EOS\space}%
\providecommand \EOS [0]{\spacefactor3000\relax}%
\providecommand \BibitemShut  [1]{\csname bibitem#1\endcsname}%
\let\auto@bib@innerbib\@empty
\bibitem [{\citenamefont {Read}\ and\ \citenamefont
  {Shockley}(1950)}]{read1950dislocation}%
  \BibitemOpen
  \bibfield  {author} {\bibinfo {author} {\bibfnamefont {W.}~\bibnamefont
  {Read}}\ and\ \bibinfo {author} {\bibfnamefont {W.}~\bibnamefont
  {Shockley}},\ }\bibfield  {title} {\bibinfo {title} {Dislocation models of
  crystal grain boundaries},\ }\href@noop {} {\bibfield  {journal} {\bibinfo
  {journal} {Physical Review}\ }\textbf {\bibinfo {volume} {78}},\ \bibinfo
  {pages} {275} (\bibinfo {year} {1950})}\BibitemShut {NoStop}%
\bibitem [{\citenamefont {Hirth}\ and\ \citenamefont
  {Lothe}(1982)}]{hirth1982theory}%
  \BibitemOpen
  \bibfield  {author} {\bibinfo {author} {\bibfnamefont {J.}~\bibnamefont
  {Hirth}}\ and\ \bibinfo {author} {\bibfnamefont {J.}~\bibnamefont {Lothe}},\
  }\href@noop {} {\emph {\bibinfo {title} {Theory of Dislocations}}}\ (\bibinfo
   {publisher} {Wiley},\ \bibinfo {year} {1982})\BibitemShut {NoStop}%
\bibitem [{\citenamefont {Du}\ and\ \citenamefont
  {Chen}(2005)}]{du2005systematic}%
  \BibitemOpen
  \bibfield  {author} {\bibinfo {author} {\bibfnamefont {Y.}~\bibnamefont
  {Du}}\ and\ \bibinfo {author} {\bibfnamefont {L.}~\bibnamefont {Chen}},\
  }\bibfield  {title} {\bibinfo {title} {Systematic analysis of grain boundary
  energy in bcc metals using the broken-bond model},\ }\href@noop {} {\bibfield
   {journal} {\bibinfo  {journal} {Acta Materialia}\ }\textbf {\bibinfo
  {volume} {53}},\ \bibinfo {pages} {2539} (\bibinfo {year}
  {2005})}\BibitemShut {NoStop}%
\bibitem [{\citenamefont {Rohrer}(2011)}]{rohrer2011grain}%
  \BibitemOpen
  \bibfield  {author} {\bibinfo {author} {\bibfnamefont {G.}~\bibnamefont
  {Rohrer}},\ }\bibfield  {title} {\bibinfo {title} {Grain boundary energy
  anisotropy: A review},\ }\href@noop {} {\bibfield  {journal} {\bibinfo
  {journal} {Journal of Materials Science}\ }\textbf {\bibinfo {volume} {46}},\
  \bibinfo {pages} {5881} (\bibinfo {year} {2011})}\BibitemShut {NoStop}%
\bibitem [{\citenamefont {Mishin}(2010)}]{mishin2010atomistic}%
  \BibitemOpen
  \bibfield  {author} {\bibinfo {author} {\bibfnamefont {Y.}~\bibnamefont
  {Mishin}},\ }\bibfield  {title} {\bibinfo {title} {Atomistic modeling of
  grain boundary structure and segregation},\ }\href@noop {} {\bibfield
  {journal} {\bibinfo  {journal} {Acta Materialia}\ }\textbf {\bibinfo {volume}
  {58}},\ \bibinfo {pages} {1117} (\bibinfo {year} {2010})}\BibitemShut
  {NoStop}%
\bibitem [{\citenamefont {Seah}(1980)}]{seah1980adsorption}%
  \BibitemOpen
  \bibfield  {author} {\bibinfo {author} {\bibfnamefont {M.}~\bibnamefont
  {Seah}},\ }\bibfield  {title} {\bibinfo {title} {Adsorption-induced interface
  decohesion},\ }\href@noop {} {\bibfield  {journal} {\bibinfo  {journal} {Acta
  Metallurgica}\ }\textbf {\bibinfo {volume} {28}},\ \bibinfo {pages} {955}
  (\bibinfo {year} {1980})}\BibitemShut {NoStop}%
\bibitem [{\citenamefont {Lejček}(2010)}]{lejcek2010grain}%
  \BibitemOpen
  \bibfield  {author} {\bibinfo {author} {\bibfnamefont {P.}~\bibnamefont
  {Lejček}},\ }\href@noop {} {\emph {\bibinfo {title} {Grain Boundary
  Segregation in Metals}}}\ (\bibinfo  {publisher} {Springer},\ \bibinfo {year}
  {2010})\BibitemShut {NoStop}%
\bibitem [{\citenamefont {McLean}(1957)}]{mclean1957grain}%
  \BibitemOpen
  \bibfield  {author} {\bibinfo {author} {\bibfnamefont {D.}~\bibnamefont
  {McLean}},\ }\bibfield  {title} {\bibinfo {title} {Grain boundary segregation
  in metals},\ }\href@noop {} {\bibfield  {journal} {\bibinfo  {journal}
  {Reports on Progress in Physics}\ }\textbf {\bibinfo {volume} {18}},\
  \bibinfo {pages} {266} (\bibinfo {year} {1957})}\BibitemShut {NoStop}%
\bibitem [{\citenamefont {Foiles}(1985)}]{foiles1985calculation}%
  \BibitemOpen
  \bibfield  {author} {\bibinfo {author} {\bibfnamefont {S.}~\bibnamefont
  {Foiles}},\ }\bibfield  {title} {\bibinfo {title} {Calculation of the
  segregation of impurities to grain boundaries in metals},\ }\href@noop {}
  {\bibfield  {journal} {\bibinfo  {journal} {Physical Review B}\ }\textbf
  {\bibinfo {volume} {32}},\ \bibinfo {pages} {7685} (\bibinfo {year}
  {1985})}\BibitemShut {NoStop}%
\bibitem [{\citenamefont {Murdoch}\ and\ \citenamefont
  {Schuh}(2013)}]{murdoch2013grain}%
  \BibitemOpen
  \bibfield  {author} {\bibinfo {author} {\bibfnamefont {H.}~\bibnamefont
  {Murdoch}}\ and\ \bibinfo {author} {\bibfnamefont {C.}~\bibnamefont
  {Schuh}},\ }\bibfield  {title} {\bibinfo {title} {Grain boundary segregation
  enthalpies and energies in binary alloys from atomistic simulations},\
  }\href@noop {} {\bibfield  {journal} {\bibinfo  {journal} {Acta Materialia}\
  }\textbf {\bibinfo {volume} {61}},\ \bibinfo {pages} {2121} (\bibinfo {year}
  {2013})}\BibitemShut {NoStop}%
\bibitem [{\citenamefont
  {Darvishi~Kamachali}(2020)}]{darvishikamachali2020model}%
  \BibitemOpen
  \bibfield  {author} {\bibinfo {author} {\bibfnamefont {R.}~\bibnamefont
  {Darvishi~Kamachali}},\ }\bibfield  {title} {\bibinfo {title} {{A model for
  grain boundary thermodynamics}},\ }\href@noop {} {\bibfield  {journal}
  {\bibinfo  {journal} {RSC Advances}\ }\textbf {\bibinfo {volume} {10}},\
  \bibinfo {pages} {26728} (\bibinfo {year} {2020})}\BibitemShut {NoStop}%
\bibitem [{\citenamefont {Darvishi~Kamachali}\ \emph
  {et~al.}(2020)\citenamefont {Darvishi~Kamachali}, \citenamefont
  {Kwiatkowski~da Silva}, \citenamefont {McEniry}, \citenamefont {Ponge},
  \citenamefont {Gault}, \citenamefont {Neugebauer},\ and\ \citenamefont
  {Raabe}}]{darvishikamachali2020segregation}%
  \BibitemOpen
  \bibfield  {author} {\bibinfo {author} {\bibfnamefont {R.}~\bibnamefont
  {Darvishi~Kamachali}}, \bibinfo {author} {\bibfnamefont {A.}~\bibnamefont
  {Kwiatkowski~da Silva}}, \bibinfo {author} {\bibfnamefont {E.}~\bibnamefont
  {McEniry}}, \bibinfo {author} {\bibfnamefont {D.}~\bibnamefont {Ponge}},
  \bibinfo {author} {\bibfnamefont {B.}~\bibnamefont {Gault}}, \bibinfo
  {author} {\bibfnamefont {J.}~\bibnamefont {Neugebauer}},\ and\ \bibinfo
  {author} {\bibfnamefont {D.}~\bibnamefont {Raabe}},\ }\bibfield  {title}
  {\bibinfo {title} {{Segregation-assisted spinodal and transient spinodal
  phase separation at grain boundaries}},\ }\href
  {https://doi.org/10.1038/s41524-020-00456-7} {\bibfield  {journal} {\bibinfo
  {journal} {npj Computational Materials}\ }\textbf {\bibinfo {volume} {6}},\
  \bibinfo {pages} {191} (\bibinfo {year} {2020})}\BibitemShut {NoStop}%
\bibitem [{\citenamefont {Wang}\ and\ \citenamefont
  {Kamachali}(2021{\natexlab{a}})}]{wang2021density}%
  \BibitemOpen
  \bibfield  {author} {\bibinfo {author} {\bibfnamefont {L.}~\bibnamefont
  {Wang}}\ and\ \bibinfo {author} {\bibfnamefont {R.~D.}\ \bibnamefont
  {Kamachali}},\ }\bibfield  {title} {\bibinfo {title} {{Density-based grain
  boundary phase diagrams: Application to Fe-Mn-Cr, Fe-Mn-Ni, Fe-Mn-Co,
  Fe-Cr-Ni and Fe-Cr-Co alloy systems}},\ }\href
  {https://doi.org/https://doi.org/10.1016/j.actamat.2021.116668} {\bibfield
  {journal} {\bibinfo  {journal} {Acta Materialia}\ ,\ \bibinfo {pages}
  {116668}} (\bibinfo {year} {2021}{\natexlab{a}})}\BibitemShut {NoStop}%
\bibitem [{\citenamefont {Zhou}\ \emph {et~al.}(2021)\citenamefont {Zhou},
  \citenamefont {Kamachali}, \citenamefont {Boyce}, \citenamefont {Clark},
  \citenamefont {Raabe},\ and\ \citenamefont {Thompson}}]{zhou2021spinodal}%
  \BibitemOpen
  \bibfield  {author} {\bibinfo {author} {\bibfnamefont {X.}~\bibnamefont
  {Zhou}}, \bibinfo {author} {\bibfnamefont {R.~D.}\ \bibnamefont {Kamachali}},
  \bibinfo {author} {\bibfnamefont {B.~L.}\ \bibnamefont {Boyce}}, \bibinfo
  {author} {\bibfnamefont {B.~G.}\ \bibnamefont {Clark}}, \bibinfo {author}
  {\bibfnamefont {D.}~\bibnamefont {Raabe}},\ and\ \bibinfo {author}
  {\bibfnamefont {G.~B.}\ \bibnamefont {Thompson}},\ }\bibfield  {title}
  {\bibinfo {title} {Spinodal decomposition in nanocrystalline alloys},\
  }\href@noop {} {\bibfield  {journal} {\bibinfo  {journal} {Acta Materialia}\
  ,\ \bibinfo {pages} {117054}} (\bibinfo {year} {2021})}\BibitemShut {NoStop}%
\bibitem [{\citenamefont {Wang}\ and\ \citenamefont
  {Kamachali}(2021{\natexlab{b}})}]{wang2021incorporating}%
  \BibitemOpen
  \bibfield  {author} {\bibinfo {author} {\bibfnamefont {L.}~\bibnamefont
  {Wang}}\ and\ \bibinfo {author} {\bibfnamefont {R.~D.}\ \bibnamefont
  {Kamachali}},\ }\bibfield  {title} {\bibinfo {title} {Incorporating
  elasticity into calphad-informed density-based grain boundary phase diagrams
  reveals segregation transition in al-cu and al-cu-mg alloys},\ }\href@noop {}
  {\bibfield  {journal} {\bibinfo  {journal} {Computational Materials Science}\
  }\textbf {\bibinfo {volume} {199}},\ \bibinfo {pages} {110717} (\bibinfo
  {year} {2021}{\natexlab{b}})}\BibitemShut {NoStop}%
\bibitem [{\citenamefont {Wang}\ and\ \citenamefont
  {Kamachali}(2023)}]{wang2023calphad}%
  \BibitemOpen
  \bibfield  {author} {\bibinfo {author} {\bibfnamefont {L.}~\bibnamefont
  {Wang}}\ and\ \bibinfo {author} {\bibfnamefont {R.~D.}\ \bibnamefont
  {Kamachali}},\ }\bibfield  {title} {\bibinfo {title} {Calphad integrated
  grain boundary co-segregation design: Towards safe high-entropy alloys},\
  }\href@noop {} {\bibfield  {journal} {\bibinfo  {journal} {Journal of Alloys
  and Compounds}\ }\textbf {\bibinfo {volume} {933}},\ \bibinfo {pages}
  {167717} (\bibinfo {year} {2023})}\BibitemShut {NoStop}%
\bibitem [{\citenamefont {Wallis}\ and\ \citenamefont
  {Kamachali}(2023)}]{wallis2023grain}%
  \BibitemOpen
  \bibfield  {author} {\bibinfo {author} {\bibfnamefont {T.}~\bibnamefont
  {Wallis}}\ and\ \bibinfo {author} {\bibfnamefont {R.~D.}\ \bibnamefont
  {Kamachali}},\ }\bibfield  {title} {\bibinfo {title} {Grain boundary
  structural variations amplify segregation transition and stabilize
  co-existing spinodal interfacial phases},\ }\href@noop {} {\bibfield
  {journal} {\bibinfo  {journal} {Acta Materialia}\ }\textbf {\bibinfo {volume}
  {242}},\ \bibinfo {pages} {118446} (\bibinfo {year} {2023})}\BibitemShut
  {NoStop}%
\bibitem [{\citenamefont {Kamachali}\ \emph {et~al.}(2024)\citenamefont
  {Kamachali}, \citenamefont {Wallis}, \citenamefont {Ikeda}, \citenamefont
  {Saikia}, \citenamefont {Ahmadian}, \citenamefont {Liebscher}, \citenamefont
  {Hickel},\ and\ \citenamefont {Maa{\ss}}}]{kamachali2024giant}%
  \BibitemOpen
  \bibfield  {author} {\bibinfo {author} {\bibfnamefont {R.~D.}\ \bibnamefont
  {Kamachali}}, \bibinfo {author} {\bibfnamefont {T.}~\bibnamefont {Wallis}},
  \bibinfo {author} {\bibfnamefont {Y.}~\bibnamefont {Ikeda}}, \bibinfo
  {author} {\bibfnamefont {U.}~\bibnamefont {Saikia}}, \bibinfo {author}
  {\bibfnamefont {A.}~\bibnamefont {Ahmadian}}, \bibinfo {author}
  {\bibfnamefont {C.~H.}\ \bibnamefont {Liebscher}}, \bibinfo {author}
  {\bibfnamefont {T.}~\bibnamefont {Hickel}},\ and\ \bibinfo {author}
  {\bibfnamefont {R.}~\bibnamefont {Maa{\ss}}},\ }\bibfield  {title} {\bibinfo
  {title} {Giant segregation transition as origin of liquid metal embrittlement
  in the fe-zn system},\ }\href@noop {} {\bibfield  {journal} {\bibinfo
  {journal} {Scripta Materialia}\ }\textbf {\bibinfo {volume} {238}},\ \bibinfo
  {pages} {115758} (\bibinfo {year} {2024})}\BibitemShut {NoStop}%
\bibitem [{\citenamefont {Ratanaphan}\ \emph {et~al.}(2015)\citenamefont
  {Ratanaphan}, \citenamefont {Olmsted}, \citenamefont {Bulatov}, \citenamefont
  {Holm}, \citenamefont {Rollett},\ and\ \citenamefont
  {Rohrer}}]{ratanaphan2015grain}%
  \BibitemOpen
  \bibfield  {author} {\bibinfo {author} {\bibfnamefont {S.}~\bibnamefont
  {Ratanaphan}}, \bibinfo {author} {\bibfnamefont {D.~L.}\ \bibnamefont
  {Olmsted}}, \bibinfo {author} {\bibfnamefont {V.~V.}\ \bibnamefont
  {Bulatov}}, \bibinfo {author} {\bibfnamefont {E.~A.}\ \bibnamefont {Holm}},
  \bibinfo {author} {\bibfnamefont {A.~D.}\ \bibnamefont {Rollett}},\ and\
  \bibinfo {author} {\bibfnamefont {G.~S.}\ \bibnamefont {Rohrer}},\ }\bibfield
   {title} {\bibinfo {title} {{Grain boundary energies in body-centered cubic
  metals}},\ }\href@noop {} {\bibfield  {journal} {\bibinfo  {journal} {Acta
  Materialia}\ }\textbf {\bibinfo {volume} {88}},\ \bibinfo {pages} {346}
  (\bibinfo {year} {2015})}\BibitemShut {NoStop}%
\bibitem [{\citenamefont {Wallis}\ and\ \citenamefont
  {Darvishi~Kamachali}(2025{\natexlab{a}})}]{wallis2025linking1}%
  \BibitemOpen
  \bibfield  {author} {\bibinfo {author} {\bibfnamefont {T.}~\bibnamefont
  {Wallis}}\ and\ \bibinfo {author} {\bibfnamefont {R.}~\bibnamefont
  {Darvishi~Kamachali}},\ }\bibfield  {title} {\bibinfo {title} {Linking
  atomistic and phase-field modelling of grain boundaries i: Coarse-graining
  atomistic structures},\ }\href@noop {} {\bibfield  {journal} {\bibinfo
  {journal} {Under Review}\ } (\bibinfo {year}
  {2025}{\natexlab{a}})}\BibitemShut {NoStop}%
\bibitem [{\citenamefont {Wallis}\ and\ \citenamefont
  {Darvishi~Kamachali}(2025{\natexlab{b}})}]{wallis2025linking2}%
  \BibitemOpen
  \bibfield  {author} {\bibinfo {author} {\bibfnamefont {T.}~\bibnamefont
  {Wallis}}\ and\ \bibinfo {author} {\bibfnamefont {R.}~\bibnamefont
  {Darvishi~Kamachali}},\ }\bibfield  {title} {\bibinfo {title} {Linking
  atomistic and phase-field modelling of grain boundaries ii: Incorporating
  atomistic potentials into free energy functional},\ }\href@noop {} {\bibfield
   {journal} {\bibinfo  {journal} {Under Review}\ } (\bibinfo {year}
  {2025}{\natexlab{b}})}\BibitemShut {NoStop}%
\bibitem [{\citenamefont {Hu}\ \emph {et~al.}(2020)\citenamefont {Hu},
  \citenamefont {Wang}, \citenamefont {Wang}, \citenamefont {Darling},
  \citenamefont {Kecskes},\ and\ \citenamefont {Liu}}]{hu2020solute}%
  \BibitemOpen
  \bibfield  {author} {\bibinfo {author} {\bibfnamefont {Y.-J.}\ \bibnamefont
  {Hu}}, \bibinfo {author} {\bibfnamefont {Y.}~\bibnamefont {Wang}}, \bibinfo
  {author} {\bibfnamefont {W.~Y.}\ \bibnamefont {Wang}}, \bibinfo {author}
  {\bibfnamefont {K.~A.}\ \bibnamefont {Darling}}, \bibinfo {author}
  {\bibfnamefont {L.~J.}\ \bibnamefont {Kecskes}},\ and\ \bibinfo {author}
  {\bibfnamefont {Z.-K.}\ \bibnamefont {Liu}},\ }\bibfield  {title} {\bibinfo
  {title} {{Solute effects on the $\Sigma$3 111 [11-0] tilt grain boundary in BCC Fe:
  Grain boundary segregation, stability, and embrittlement}},\ }\href@noop {}
  {\bibfield  {journal} {\bibinfo  {journal} {Computational Materials Science}\
  }\textbf {\bibinfo {volume} {171}},\ \bibinfo {pages} {109271} (\bibinfo
  {year} {2020})}\BibitemShut {NoStop}%
\bibitem [{\citenamefont {Frolov}\ and\ \citenamefont
  {Mishin}(2012{\natexlab{a}})}]{frolov2012thermodynamicsI}%
  \BibitemOpen
  \bibfield  {author} {\bibinfo {author} {\bibfnamefont {T.}~\bibnamefont
  {Frolov}}\ and\ \bibinfo {author} {\bibfnamefont {Y.}~\bibnamefont
  {Mishin}},\ }\bibfield  {title} {\bibinfo {title} {Thermodynamics of coherent
  interfaces under mechanical stresses. i. theory},\ }\href@noop {} {\bibfield
  {journal} {\bibinfo  {journal} {Physical Review B—Condensed Matter and
  Materials Physics}\ }\textbf {\bibinfo {volume} {85}},\ \bibinfo {pages}
  {224106} (\bibinfo {year} {2012}{\natexlab{a}})}\BibitemShut {NoStop}%
\bibitem [{\citenamefont {Frolov}\ and\ \citenamefont
  {Mishin}(2012{\natexlab{b}})}]{frolov2012thermodynamicsII}%
  \BibitemOpen
  \bibfield  {author} {\bibinfo {author} {\bibfnamefont {T.}~\bibnamefont
  {Frolov}}\ and\ \bibinfo {author} {\bibfnamefont {Y.}~\bibnamefont
  {Mishin}},\ }\bibfield  {title} {\bibinfo {title} {Thermodynamics of coherent
  interfaces under mechanical stresses. ii. application to atomistic simulation
  of grain boundaries},\ }\href@noop {} {\bibfield  {journal} {\bibinfo
  {journal} {Physical Review B—Condensed Matter and Materials Physics}\
  }\textbf {\bibinfo {volume} {85}},\ \bibinfo {pages} {224107} (\bibinfo
  {year} {2012}{\natexlab{b}})}\BibitemShut {NoStop}%
\bibitem [{\citenamefont {Dehm}\ and\ \citenamefont
  {Cairney}(2022)}]{dehm2022implication}%
  \BibitemOpen
  \bibfield  {author} {\bibinfo {author} {\bibfnamefont {G.}~\bibnamefont
  {Dehm}}\ and\ \bibinfo {author} {\bibfnamefont {J.}~\bibnamefont {Cairney}},\
  }\bibfield  {title} {\bibinfo {title} {Implication of grain-boundary
  structure and chemistry on plasticity and failure},\ }\href@noop {}
  {\bibfield  {journal} {\bibinfo  {journal} {MRS Bulletin}\ }\textbf {\bibinfo
  {volume} {47}},\ \bibinfo {pages} {800} (\bibinfo {year} {2022})}\BibitemShut
  {NoStop}%
\bibitem [{\citenamefont {Bishara}\ \emph {et~al.}(2020)\citenamefont
  {Bishara}, \citenamefont {Ghidelli},\ and\ \citenamefont
  {Dehm}}]{bishara2020approaches}%
  \BibitemOpen
  \bibfield  {author} {\bibinfo {author} {\bibfnamefont {H.}~\bibnamefont
  {Bishara}}, \bibinfo {author} {\bibfnamefont {M.}~\bibnamefont {Ghidelli}},\
  and\ \bibinfo {author} {\bibfnamefont {G.}~\bibnamefont {Dehm}},\ }\bibfield
  {title} {\bibinfo {title} {Approaches to measure the resistivity of grain
  boundaries in metals with high sensitivity and spatial resolution: A case
  study employing cu},\ }\href@noop {} {\bibfield  {journal} {\bibinfo
  {journal} {ACS Applied Electronic Materials}\ }\textbf {\bibinfo {volume}
  {2}},\ \bibinfo {pages} {2049} (\bibinfo {year} {2020})}\BibitemShut
  {NoStop}%
\bibitem [{\citenamefont {R{\"o}sner}\ \emph {et~al.}(2014)\citenamefont
  {R{\"o}sner}, \citenamefont {Peterlechner}, \citenamefont {K{\"u}bel},
  \citenamefont {Schmidt},\ and\ \citenamefont
  {Wilde}}]{Rosner2014Ultramicroscopy}%
  \BibitemOpen
  \bibfield  {author} {\bibinfo {author} {\bibfnamefont {H.}~\bibnamefont
  {R{\"o}sner}}, \bibinfo {author} {\bibfnamefont {M.}~\bibnamefont
  {Peterlechner}}, \bibinfo {author} {\bibfnamefont {C.}~\bibnamefont
  {K{\"u}bel}}, \bibinfo {author} {\bibfnamefont {V.}~\bibnamefont {Schmidt}},\
  and\ \bibinfo {author} {\bibfnamefont {G.}~\bibnamefont {Wilde}},\ }\bibfield
   {title} {\bibinfo {title} {Density changes in shear bands of a metallic
  glass determined by correlative analytical transmission electron
  microscopy},\ }\href {https://doi.org/10.1016/j.ultramic.2014.03.006}
  {\bibfield  {journal} {\bibinfo  {journal} {Ultramicroscopy}\ }\textbf
  {\bibinfo {volume} {142}},\ \bibinfo {pages} {1} (\bibinfo {year}
  {2014})}\BibitemShut {NoStop}%
\bibitem [{\citenamefont {Schmidt}\ \emph {et~al.}(2015)\citenamefont
  {Schmidt}, \citenamefont {R{\"o}sner}, \citenamefont {Peterlechner},
  \citenamefont {Wilde},\ and\ \citenamefont {Voyles}}]{Schmidt2015PRL}%
  \BibitemOpen
  \bibfield  {author} {\bibinfo {author} {\bibfnamefont {V.}~\bibnamefont
  {Schmidt}}, \bibinfo {author} {\bibfnamefont {H.}~\bibnamefont {R{\"o}sner}},
  \bibinfo {author} {\bibfnamefont {M.}~\bibnamefont {Peterlechner}}, \bibinfo
  {author} {\bibfnamefont {G.}~\bibnamefont {Wilde}},\ and\ \bibinfo {author}
  {\bibfnamefont {P.~M.}\ \bibnamefont {Voyles}},\ }\bibfield  {title}
  {\bibinfo {title} {Quantitative measurement of density in a shear band of
  metallic glass monitored along its propagation direction},\ }\href
  {https://doi.org/10.1103/PhysRevLett.115.035501} {\bibfield  {journal}
  {\bibinfo  {journal} {Phys. Rev. Lett.}\ }\textbf {\bibinfo {volume} {115}},\
  \bibinfo {pages} {035501} (\bibinfo {year} {2015})}\BibitemShut {NoStop}%
\bibitem [{\citenamefont {Buranova}\ \emph {et~al.}(2016)\citenamefont
  {Buranova}, \citenamefont {Rösner}, \citenamefont {Divinski}, \citenamefont
  {Imlau},\ and\ \citenamefont {Wilde}}]{buranova2016}%
  \BibitemOpen
  \bibfield  {author} {\bibinfo {author} {\bibfnamefont {Y.}~\bibnamefont
  {Buranova}}, \bibinfo {author} {\bibfnamefont {H.}~\bibnamefont {Rösner}},
  \bibinfo {author} {\bibfnamefont {S.~V.}\ \bibnamefont {Divinski}}, \bibinfo
  {author} {\bibfnamefont {R.}~\bibnamefont {Imlau}},\ and\ \bibinfo {author}
  {\bibfnamefont {G.}~\bibnamefont {Wilde}},\ }\bibfield  {title} {\bibinfo
  {title} {Quantitative measurements of grain boundary excess volume from
  haadf–stem micrographs},\ }\href
  {https://doi.org/10.1016/j.actamat.2016.01.016} {\bibfield  {journal}
  {\bibinfo  {journal} {Acta Materialia}\ }\textbf {\bibinfo {volume} {106}},\
  \bibinfo {pages} {367} (\bibinfo {year} {2016})}\BibitemShut {NoStop}%
\bibitem [{\citenamefont {Cheng}\ \emph {et~al.}(2021)\citenamefont {Cheng},
  \citenamefont {Yin}, \citenamefont {Grutzik}, \citenamefont {Zinkle},
  \citenamefont {Bei}, \citenamefont {Xia},\ and\ \citenamefont
  {Li}}]{cheng2021acsNano}%
  \BibitemOpen
  \bibfield  {author} {\bibinfo {author} {\bibfnamefont {G.}~\bibnamefont
  {Cheng}}, \bibinfo {author} {\bibfnamefont {K.}~\bibnamefont {Yin}}, \bibinfo
  {author} {\bibfnamefont {S.~J.}\ \bibnamefont {Grutzik}}, \bibinfo {author}
  {\bibfnamefont {S.~J.}\ \bibnamefont {Zinkle}}, \bibinfo {author}
  {\bibfnamefont {H.}~\bibnamefont {Bei}}, \bibinfo {author} {\bibfnamefont
  {S.}~\bibnamefont {Xia}},\ and\ \bibinfo {author} {\bibfnamefont
  {J.}~\bibnamefont {Li}},\ }\bibfield  {title} {\bibinfo {title} {Mapping free
  volume distributions in oxide glasses by four-dimensional scanning
  transmission electron microscopy},\ }\href
  {https://doi.org/10.1021/acsnano.1c06367} {\bibfield  {journal} {\bibinfo
  {journal} {ACS Nano}\ }\textbf {\bibinfo {volume} {15}},\ \bibinfo {pages}
  {19435} (\bibinfo {year} {2021})}\BibitemShut {NoStop}%
\bibitem [{\citenamefont {Bokeloh}\ \emph {et~al.}(2011)\citenamefont
  {Bokeloh}, \citenamefont {Divinski}, \citenamefont {Reglitz},\ and\
  \citenamefont {Wilde}}]{Bokeloh2011PRL}%
  \BibitemOpen
  \bibfield  {author} {\bibinfo {author} {\bibfnamefont {J.}~\bibnamefont
  {Bokeloh}}, \bibinfo {author} {\bibfnamefont {S.~V.}\ \bibnamefont
  {Divinski}}, \bibinfo {author} {\bibfnamefont {G.}~\bibnamefont {Reglitz}},\
  and\ \bibinfo {author} {\bibfnamefont {G.}~\bibnamefont {Wilde}},\ }\bibfield
   {title} {\bibinfo {title} {Tracer measurements of atomic diffusion inside
  shear bands of a bulk metallic glass},\ }\href
  {https://doi.org/10.1103/PhysRevLett.107.235503} {\bibfield  {journal}
  {\bibinfo  {journal} {Phys. Rev. Lett.}\ }\textbf {\bibinfo {volume} {107}},\
  \bibinfo {pages} {235503} (\bibinfo {year} {2011})}\BibitemShut {NoStop}%
\end{thebibliography}
%


\section*{Acknowledgments}
We acknowledge the financial support from the German research foundation (DFG) within the project \emph{DA 1655/3-1}, \emph{DA 1655/4-1} and the Heisenberg programme project \emph{DA 1655/2-1}.

\section*{Statement}
The current version of this manuscript is released for the sake of discussion and possible feedback.
Further data and codes of this study will be made soon available.
Methods and additional discussions are under preparation to be presented in a Supplementary Material (SM) document to be attached in the next version of the manuscript. 





\end{document}